\documentclass[namedreferences]{SolarPhysics}
\usepackage[optionalrh]{spr-sola-addons} 
\usepackage{graphicx}        
\usepackage{color}           
\usepackage{url}             

\newcommand{\etal}{{\it et al.}}
\newcommand{\eg}{{\it e.g. }}

\begin{document}

\begin{article}

\begin{opening}

\title{Surface-focused Seismic Holography of Sunspots:  I. Observations}

\author{D.C.~\surname{Braun}$^{1}$\sep
        A.C.~\surname{Birch}$^{1}$
       }
\runningauthor{D. Braun and A. Birch}
\runningtitle{Holography of Sunspots: Observations}

   \institute{$^{1}$ NWRA, CoRA Division, Boulder, Colorado, USA;
                     email: \url{dbraun@cora.nwra.com} email: \url{aaronb@cora.nwra.com}\\ 
             }

\begin{abstract}
We present a comprehensive set of observations of the interaction
of $p$-mode oscillations with sunspots using surface-focused seismic
holography. Maps of travel-time shifts, relative to
quiet-Sun travel times, are shown for incoming and outgoing
$p$ modes as well as their mean and difference.  We compare results using
phase-speed filters with results obtained with 
filters that isolate single $p$-mode ridges,
and further divide the data into multiple temporal frequency bandpasses.
The $f$ mode is removed from the data.  The variations of the resulting
travel-time shifts with magnetic-field strength and with the
filter parameters are explored. We find that spatial averages of
these shifts within sunspot umbrae, penumbrae, and surrounding
plage often show strong frequency variations at fixed phase speed.
In addition, we find that positive values of the mean and difference
travel-time shifts appear exclusively in waves observed with
phase-speed filters that are dominated by power in the 
low-frequency wing of the $p_1$ ridge.
We assess the ratio of incoming to outgoing $p$-mode power
using the ridge filters and compare surface-focused holography measurements
with the results of earlier published $p$-mode scattering
measurements using Fourier-Hankel decomposition.
\end{abstract}
\keywords{Active Regions, Magnetic Fields, Helioseismology, Observations}
\end{opening}

\section{Introduction}
     \label{S-Introduction} 

The use of solar acoustic ($p$-mode) waves to probe the
subsurface structure of active regions (ARs) was first proposed
by \inlinecite{Thomas1982}. While efforts have been 
made to deduce properties of sunspots from interpretations of oscillations 
observed within sunspots (see reviews by 
\opencite{Lites1992}, and \opencite{Bogdan2000}), recent advances in
sunspot seismology have been largely driven 
by the observations of strong influences of
sunspots (and ARs in general) on externally impinging
$p$ modes. This includes both 
absorption (\eg \opencite{Braun1988}; \opencite{Bogdan1993})
and changes in phase (often characterized in terms of a
change in travel time: \eg \opencite{Braun1992}; \opencite{Braun1995};
\opencite{Duvall1996}).
A prevalent,
largely phenomenological, approach to exploiting the travel-time 
shifts (relative to travel times in the quiet Sun) to model the 
subsurface properties of sunspots has been 
the characterization of the spot as a perturbation
in the background sound speed. These types of models have
been constructed using observations from 
a variety of local-helioseismic techniques, including
Fourier-Hankel decomposition (\eg \opencite{Fan1995}), time-distance 
(\eg \opencite{Kosovichev1996}; \opencite{Kosovichev2000};
\opencite{Jensen2001}),
ring-diagrams (\eg \opencite{Basu2004}), and holography 
(\eg \opencite{Lindsey2005b}b).
The discovery of travel-time asymmetries between waves propagating
towards and away from sunspots \cite{Duvall1996} have
lead to the inclusion of subsurface flows in many of these
efforts. Travel times inferred from time-distance (TD) helioseismology
have in particular been inverted to model flows and sound-speed 
perturbations using a variety of assumptions including
Fermat's Principle and the ray approximation
(\eg \opencite{Kosovichev1997}; \opencite{Kosovichev2000};
\opencite{Zhao2001}; \opencite{Hughes2005}),
the Fresnel-Zone approximation (\eg \opencite{Jensen2001};
\opencite{Couvidat2004}),
and the Born approximation \cite{Couvidat2006}.
A consensus of many of these 3D inversions has emerged consisting
of a relatively shallow (approximately 3 Mm deep) ``slower'' sound-speed
perturbation above a ``faster'' sound-speed layer extending 10 Mm or
more below the photosphere (see review by \opencite{Gizon2005}).

These phenomenological models have been 
useful as foundations for developing 
both forward and inverse methods 
under a variety of approximations and assumptions \cite{Gizon2005}. 
At the same time, uncertainties
about the degree to which the magnetic fields may contribute (in
ways other than through associated thermal perturbations and
flows) to phase or travel-time shifts, particular in the near-surface layers, 
have persisted.  
Most local-helioseismic models of travel-time shifts, to date, do not
include provisions for contributions from unresolved near-surface layers
(near the photosphere, the typical vertical 
resolution provided by observed $p$ modes 
is around 1 Mm, \eg \opencite{Couvidat2006}). 
Notable exceptions include some 1D (horizontally
invariant) structural inversions using
ring-diagram analyses (\eg \opencite{Simmons2003}; \opencite{Basu2004}).

Some observations and inferred sound-speed models may show
direct evidence of strong near-surface contributions to the
helioseismic signatures associated with ARs. An
early example of this is the predominantly
near-surface sound-speed perturbation 
consistent with Fourier-Hankel analysis \cite{Fan1995}.
\citeauthor{Birch2007} (\citeyear{Birch2007}; Paper 2) 
examine the relevance of this 
particular result to a more recent modeling effort.
\citeauthor{Lindsey2005b} (\citeyear{Lindsey2005b}b) have shown that
helioseismic signatures beneath ARs obtained
using holography largely vanish when a surface 
(``showerglass'') phase shift, empirically related to 
photospheric magnetic flux density, is removed from the data.
Some peculiar properties of the 3D time-distance inversions 
have also been presented as evidence for surface ``contamination.''
\inlinecite{Korzennik2006} demonstrated that an inferred subsurface
sound-speed ``plume'' structure depends critically on the 
inclusion of observations made
within a sunspot penumbra and umbra. A test
of inversions for flows performed by masking only the umbra 
showed little effect of the mask \cite{Zhao2003}.
Ringlike regions of enhanced sound-speed in TD 
inversions of sunspots have also been examined as possible 
artifacts arising from the surface \cite{Couvidat2007}.
Surface effects in magnetic fields also include changes in
the upper turning points (\eg \opencite{Kosovichev1997}; 
\opencite{Braun2000}; \opencite{Barnes2001}). The observed reduction of
$p$-mode amplitudes in spots has been shown to contribute to
travel-time shifts independent of actual structural changes
\cite{Rajaguru2006} as has reduced wave excitation
(\opencite{Hanasoge2007c}; \opencite{Parchevsky2007}).

\inlinecite{Schunker2005} 
and \inlinecite{Schunker2007} found that travel-time shifts
obtained from seismic holography in sunspot penumbrae
vary with the line-of-sight angle projected
onto the plane containing the magnetic field and the vertical direction.
A similar effect has also been noted by \inlinecite{Zhao2006} with 
time-distance measurements. 
A satisfactory theory explaining these observations remains to be constructed, 
but some preliminary suggestions include 
mode conversion \cite{Schunker2006a} or radiative transfer
effects in combination with mode propagation asymmetries 
\cite{Rajaguru2007}.
Whatever the cause, the observed line-of-sight dependence of
travel-time shifts 
implies that a significant component of the shifts, at least
in sunspot penumbrae, must be photospheric in origin. 

In 1D inversions in global helioseismology (\eg \opencite{JCD1988}) and
ring-diagram analyses (\eg \opencite{Basu2004}), surface effects
are largely characterized by their frequency-dependent
contribution to the helioseismic signatures (mode or ridge
frequencies). In contrast, the observations used in 3D travel-time 
inversions are typically
made over a single, wide, frequency bandpass, and do not
easily allow the assessment of possible frequency-dependent surface terms.
However, there is increasing evidence for frequency variations
in the travel-time shifts observed in ARs 
(\opencite{Braun2000}; \opencite{Chou2000}; \opencite{Lindsey2004a}b;
 \opencite{Braun2006}; \opencite{Couvidat2007}).
\inlinecite{Braun2006} found evidence
for a frequency variation, at fixed phase
speed, of the travel times measured in active regions using
helioseismic holography. This variation exceeds the smaller
frequency variation expected from travel-time shifts computed
from a proxy sound-speed
model, with properties similar to recent two-component 
3D inversions, using the ray approximation.

The observed travel-time asymmetries in sunspots (differences in 
travel times between the incoming and outgoing 
propagating waves) have been
interpreted and modeled as due to flows 
\cite{Duvall1996,Zhao2001,Zhao2003}. The shallow inflows, within the
first 3 Mm below the surface, characteristic of
some of these $p$-mode-based TD models appear to
be inconsistent with outflows inferred 
from other methods including $f$-mode time distance
\cite{Gizon2000} and holography \cite{Braun2004}.
Some questions have been raised whether
travel-time asymmetries may arise from other mechanisms,
including the suppression of acoustic sources 
\cite{Gizon2002,Hanasoge2007c} or absorption \cite{Woodard1997,Lindsey2007}.

Including magnetic fields in helioseismic models of sunspots
appears to be a substantially more formidable task than constructing
models that include only thermal perturbations. Some progress
has been made with MHD models of the absorption in sunspots observed from 
Fourier-Hankel decomposition (\eg see the review by 
\opencite{Bogdan1995}). More recent efforts have
also addressed the observed phase shifts (\eg
\opencite{Cally2003}; \opencite{Crouch2005b}; 
\opencite{Gordovskyy2007}).  It is expected that considerable advances 
in modeling helioseismic data will follow from the current
development and application of hydrodynamic (HD) and magnetohydrodynamic
(MHD) simulations (\eg \opencite{Jensen2003b}; \opencite{Tong2003};
\opencite{Mansour2004}; \opencite{Werne2004}; \opencite{Benson2006}; 
\opencite{Hanasoge2007a}; \opencite{Hanasoge2007b}; 
\opencite{Khomenko2006}; \opencite{Shelyag2006}; \citeyear{Shelyag2007}
\opencite{Cameron2007a}; \opencite{Cameron2007b}).

Our primary motivation in this paper is to expand the 
measurements of \inlinecite{Braun2006}. We hope that
a comprehensive exposition of helioseismic observations of 
travel-time shifts, and their dependence on $p$-mode properties,
will promote and support improved modeling efforts, including
the use of numerical simulations. As we are specifically interested
in the importance of near-surface effects we also examine the
relationship between the observed travel-time shifts and
the photospheric magnetic field. Of particular importance
is the measurement of frequency variations of the travel-time
shifts, using methods similar to \inlinecite{Braun2006}.
However, we extend those measurements to include both mean
travel-time shifts and travel-time asymmetries, and
we determine spatial averages of these quantities over 
sunspot umbrae, penumbrae, and other magnetic regions. 
Our principle tool is surface-focused helioseismic 
holography (\eg \opencite{Braun2000}; \opencite{Braun2006}), 
for which the travel-time shifts are expected
to have the most sensitivity to near-surface perturbations.
The use of surface-focus holography (described
in Section~\ref{S-Analysis}) contrasts this work with other 
recent (``lateral-vantage'') 
holographic studies of ARs 
(\eg \opencite{Lindsey2005a}a; \citeyear{Lindsey2005b}b). 
In addition, to ensure a meaningful 
comparisons of our results (described in 
Section~\ref{S-Observations}) to TD observations
we use narrow annular pupils and corresponding phase-speed
filters as discussed in Section~\ref{S-phasespeed}. 
An overriding theme in our findings 
is a strong sensitivity of the results to
the choice of filter and frequency bandwidth employed.
To investigate this further, we also employ filters
centered on the $p$-mode ridges (Section~\ref{S-ridge}).
The ridge-based filters allow a detailed comparison
of surface-focused holography measurements of both travel-time shifts and
absorption with published results of Fourier-Hankel
analysis (Section~\ref{S-hankel}).

\section{Analysis} 
      \label{S-Analysis}      

Helioseismic holography (HH) is a method based on
the phase-coherent imaging of the solar interior acoustic field.
It computationally extrapolates the surface acoustic field
into the solar interior (\opencite{Lindsey1997}; \citeyear{Lindsey2000a})
to estimate the amplitudes of the waves propagating into
and out of a focus point at a chosen depth and position in the solar
interior. These amplitudes, called the ingression, ($H_-$), and
egression, ($H_+$) are estimated by a convolution of the surface 
oscillation signal ($\psi$)
(typically the line-of-sight component of
velocity observed from Dopplergrams)
with appropriate Green's functions \cite{Lindsey2000a}.
For this work, the Green's functions are computed
in the eikonal formulation (\opencite{Lindsey1997}; \citeyear{Lindsey2000a}).
For surface-focused HH, the Green's
functions represent propagators that evolve the acoustic
field forward or backward in time from a position on the solar surface into the
solar interior, and back up to the surface focus. To select
a particular set of $p$ modes, these functions are 
evaluated for over a chosen annular pupil.  A dispersion 
correction, empirically determined from statistics obtained
from measurements in the quiet Sun, is applied to the computation of the
Green's functions (see \opencite{Lindsey2000a}).

The basis of our analysis consists
of what are termed {\it local control correlations}
(\opencite{Lindsey2004b}a; \citeyear{Lindsey2005a}a).
These are directly comparable to center-annulus TD
correlations (\eg \opencite{Duvall1996}; \opencite{Braun1997}).
In the space-frequency domain, the correlation,
\begin{equation}
C_+( \textbf{r}) =  \langle {H}_{+}(\textbf{r}, \nu) 
\psi^* (\textbf{r}, \nu)\rangle_{\Delta \nu},
\label{Eq-outcorr}
\end{equation}
describes the egression control correlation, while
\begin{equation}
C_-( \textbf{r}) =  \langle \psi (\textbf{r}, \nu) 
{H}_{-}^*(\textbf{r}, \nu)  \rangle_{\Delta \nu},
\label{Eq-incorr}
\end{equation}
describes the ingression control correlation.
Here, $\psi (\textbf{r}, \nu)$ represents the temporal Fourier
transform of the surface wave field,
$\nu$ is the temporal frequency, $\textbf{r}$ is
the horizontal position on the solar surface,
and $H_{-} (\textbf{r}, \nu)$ and $H_{+} (\textbf{r}, \nu)$ represent the 
temporal Fourier transforms of the ingression and egression respectively.
The asterisk denotes complex conjugation, and the brackets indicate
an average over a chosen positive frequency range  $\Delta \nu$.

The primary quantities of interest are the travel-time shifts
which are related to the phase of the correlations,
\begin{equation}
\delta \tau_{\pm} = \arg [ C_{\pm} (\textbf{r}) ] / 2\pi{\nu}_0,
\label{Eq-corr_phase}
\end{equation}
where ${\nu}_0$ is the central frequency of the bandpass $\Delta \nu$.
These represent travel-time shifts of the observed incoming 
($\tau_{-}$) or outgoing ($\tau_{+}$) waves, as sampled
by a chosen filter, relative to the 
travel times expected for
the same ensemble of waves propagating in the solar model used
to compute the Green's functions. Small systematic
deviations of the quiet-Sun values from zero, which
vary with pupil size and filter and are likely
caused by imperfections in the Green's functions and
dispersion correction are removed by subtracting 
averaged quiet-Sun values from the observed control correlation 
phases.  Of interest are
the mean travel-time shift: 
$\delta \tau_{\rm{mean}} = (\delta \tau_+ + \delta \tau_-)/2$,
and the difference (or travel-time asymmetry):
$\delta \tau_{\rm{diff}} = \delta \tau_+ - \delta \tau_-$.

A 27 hour sequence of full disk Dopplergrams with one minute
cadence, obtained from the Michelson Doppler Imager (MDI;
\opencite{Scherrer1995}) onboard the {\it Solar and
Heliospheric Observatory} (SOHO), were used in this study.
The data set starts on 1 April 2002, 21:01 UT, and includes several
sunspot groups (NOAA groups 9885, 9886, 9887, and 9888) within
a $60 \times 60^\circ$  Postel-projected region. This area
was tracked at the Carrington rotation rate and includes four sunspots
with penumbral radii greater than 15 Mm as well as other
smaller spots. The three largest sunspots are very
similar in size, with mean umbral and penumbral radii of
7 and 18 Mm respectively.

The following steps summarize the general data reduction:
1) projection of the desired region from full-disk  Dopplergrams
to a Postel projection that rotates with a fixed Carrington rate,
2) temporal detrending by subtraction of a linear
fit to each pixel signal in time, 3) removal of poor
quality images, identified by a five-sigma deviation of
any pixel from the linear trend , 4) Fourier transform of
the data in time, 
5) (optional) correction
for the amplitude suppression in 
magnetic regions \cite{Rajaguru2006},
6) spatial Fourier transform of the data and
multiplication by a chosen filter,
7) extraction of the desired frequency
bandpass, 8) computation of Green's
functions over the appropriate pupil, 9) computation
of ingression and egression amplitudes by a 3D convolution
of the data with the Green's functions, and
10) computation of the travel-time shift maps by 
Equations~(\ref{Eq-outcorr})\,--\,(\ref{Eq-corr_phase}).

The optional correction
for amplitude suppression (step 5) involves dividing the
amplitude of each pixel in the data by
its root-mean-square value over the frequency bandpass.
In step 6 we have used two sets of filters:
phase-speed filters and ridge filters. Their description and the
results obtained from each set are described in 
Section~\ref{S-phasespeed} and Section~\ref{S-ridge}. 
For the phase-speed filters (Section~\ref{S-phasespeed}) we compare 
results with and without
the amplitude-suppression correction. 
The difference is relatively small and, 
for the ridge-filtering (Section~\ref{S-ridge}), 
we use only uncorrected data.

\section{Observations} 
      \label{S-Observations}      

\subsection{Phase-Speed Filters} 
  \label{S-phasespeed}

The phase-speed filters used are of the type specified by
\inlinecite{Couvidat2006}, namely, the three-dimensional 
Fourier transform of the
data (step 6) is multiplied by a function
\begin{equation}
F(\textbf{k}, \nu) =  \exp
\left[ - (2 \pi \nu / |\textbf{k}| - w )^2 /  2 \delta  w^2 \right]
\label{phsp_filter}
\end{equation}
where $w$ and $\delta  w$ are the mean phase speed and filter width
respectively.  We use the same set of ten filters (denoted
A through J) of \inlinecite{Braun2006}
with parameters listed in Table 1 of that paper. These filters
are of the same type, but are somewhat narrower in width, than
the eleven common filters often employed in TD
analyses (\eg \opencite{Couvidat2006}, \opencite{Zhao2006}).
Each filter is used with a corresponding pupil, over which the
ingression and egression are evaluated. This pupil is a complete annulus
defined so  that acoustic rays at a frequency of $\nu = $ 3.5 mHz
reaching the inner and outer radii span the
full width at half maximum (FWHM) of the squared filter. The
parameter $\delta w$ is related to the FWHM by 
$\delta  w = \rm{FWHM}/[2(\ln 2)^{1/2} ]$.
The filters were chosen such that the sets of FWHM and corresponding
pupils span a continuous range of phase-speed and radius respectively.

All of the phase speed filters used also remove the contribution of
the $f$ mode, a practice first used by 
\inlinecite{Giles2000}. 
Our $f$-mode cutoff consists of a high-pass filter with Gaussian roll-off
in temporal frequency. The position and rate of
the roll-off varies with spatial wavenumber
such that full transmission occurs at a frequency midway
between the $f$-mode and $p_1$ ridges and $10^{-7}$ transmission occurs at the
frequency of the $f$-mode.

The filters are applied to, and the travel-time shift maps computed from,
portions of the data extracted in 1-mHz-wide frequency bandpasses
centered at 2, 3, 4, and 5 mHz. We also compute travel-time 
shifts over a wider bandpass (2.5\,--\,5.5 mHz) typical of common
TD measurements. The use of narrow frequency bandpasses,
although frequently employed in HH, differs from typical applications of
TD correlations. Consequently the wide bandpass measurements
provide a useful check and basis for comparison. It should be noted, 
however, that
the power spectra of solar acoustic oscillations are naturally limited in
bandwidth. Power spectra computed using typical phase-speed filters
applied to MDI data, without any additional temporal filtering, show 
a concentration of between 60 and 90\% of the power within a 1 mHz 
bandwidth, depending on the choice of phase-speed filter. 

\begin{figure}    
\centerline{\includegraphics[width=\textwidth,clip=]{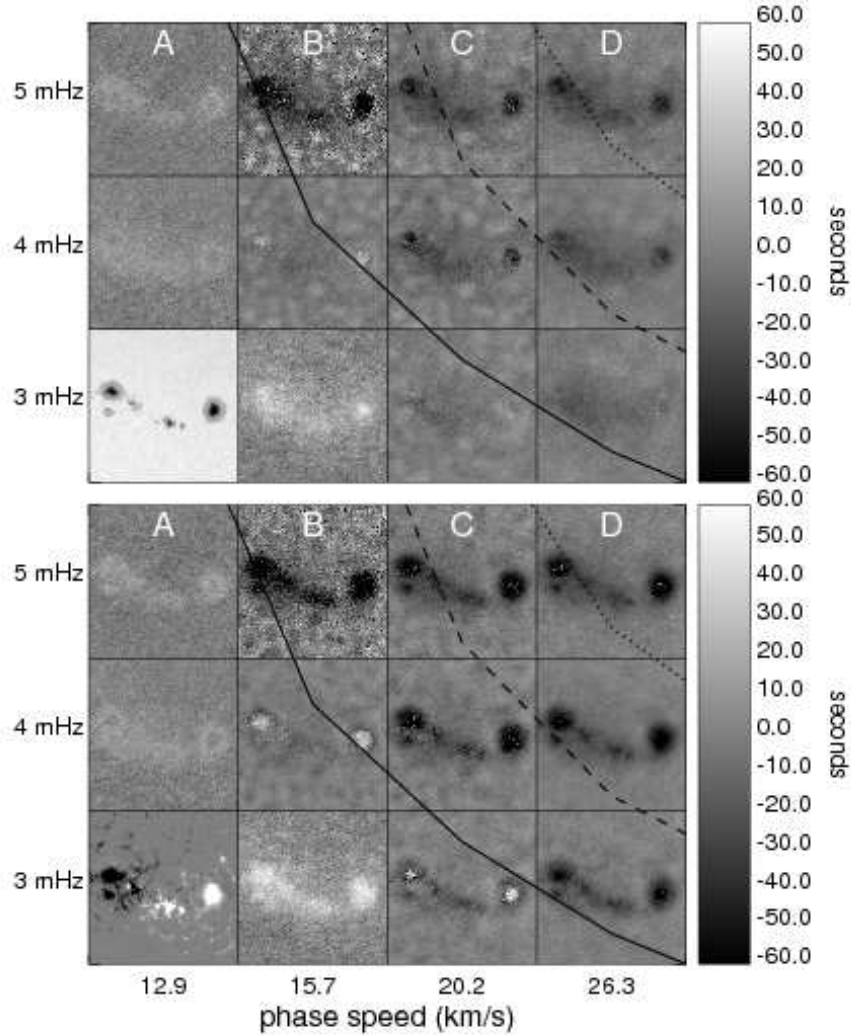}}
\caption{Maps of travel-time shifts $\delta \tau_{\rm{-}}$ (top panels) and $\delta \tau_{\rm{+}}$ 
(bottom panels) covering a portion of the region studied and showing sunspot group 9885.
The columns of maps labeled A through D indicate the phase-speed filter used,
while the rows indicate the frequency bandpass. The solid jagged line
running diagonally through the panels connects the location of
the $p_1$ ridge in the $\nu$-$w$  domain for each filter, with
the centers of the maps assigned to values of frequency and phase
speed as indicated
on the left and bottom edges of the plot. The dashed and dotted lines
indicate the locations of the $p_2$ and $p_3$ ridges respectively.
The map in the lowest-left position of the top set of panels shows a MDI 
continuum
intensity image while the map in the same position in the bottom set
shows a line-of-sight magnetogram
}
\label{stack_inout}
\end{figure}

\begin{figure}    
\centerline{\includegraphics[width=\textwidth,clip=]{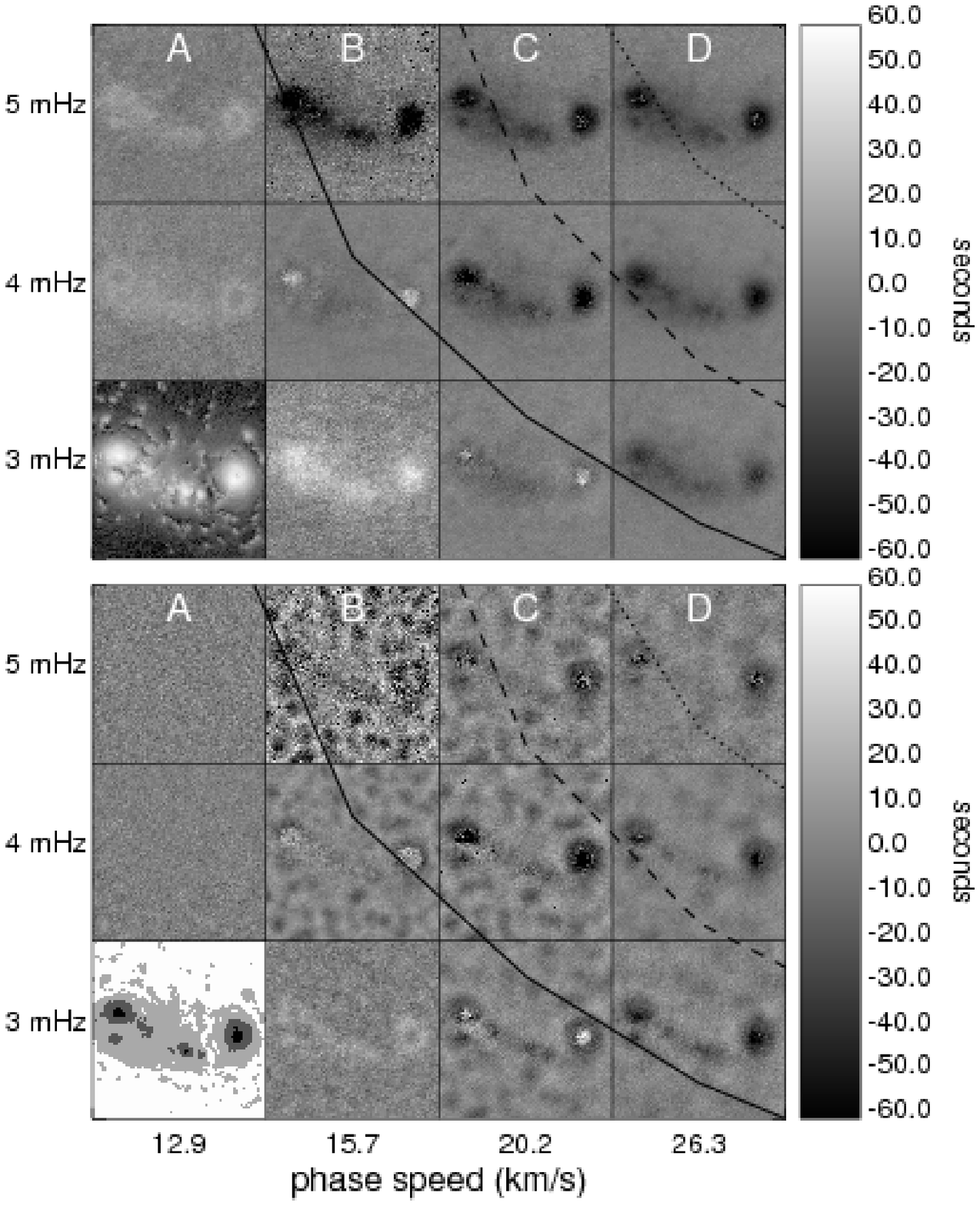}}
\caption{Maps of the travel-time shifts $\delta \tau_{\rm{mean}}$ 
(top panels) and $\delta \tau_{\rm{diff}}$ covering a portion of the region studied
and showing sunspot group 9885.
The columns of maps labeled A through D indicate the phase-speed filter used,
while the rows indicate the frequency bandpass. The solid jagged line
running diagonally through the panels connects the location of
the $p_1$ ridge in the $\nu$-$w$  domain for each filter, with
the centers of the maps assigned to values of frequency and phase
speed as indicated
on the left and bottom edges of the plot. The dashed and dotted lines
indicate the locations of the $p_2$ and $p_3$ ridges respectively.
The map in the lowest-left position of the top 
set of panels shows $B_{\rm tot}$ (see text)    
while the image in the same position in the bottom set
isolates in shades of grey different portions of the
AR for study (see text).
}
\label{stack_meandiff}
\end{figure}

Some maps of travel-time shifts computed with phase-speed
filters are shown in Figures~\ref{stack_inout} and 
\ref{stack_meandiff}. Figure~\ref{stack_inout} shows maps of 
$\delta \tau_{\rm{-}}$ (top panels) and $\delta \tau_{\rm{+}}$ (bottom panels),
while Figure~\ref{stack_meandiff} shows the corresponding maps of
$\delta \tau_{\rm{mean}}$ (top) and $\delta \tau_{\rm{diff}}$ (bottom). The
maps are stacked into columns of increasing phase-speed (left
to right) and rows representing increasing frequency (bottom
to top). For simplicity we will refer to each map by
a number-letter combination denoting filter and frequency
combination; \eg ``4B'' refers to filter B applied to 
the frequency bandpass centered at 4 mHz. 
Because of the filter masking the $f$ mode, and the decrease
in acoustic wave amplitudes at frequencies below the $p_1$ ridge,
only a subset of possible frequency-filter combinations produce
meaningful correlations. The maps analyzed here are limited to 
filters 2D\,--\,2J, 3B\,--\,3J, 4A\,--\,4J, 
and 5A\,--\,5J. 

Individual maps exhibit spatial relationships between the
travel-time shifts and the surface magnetic flux 
explored below. 
The stacking of maps in Figures~\ref{stack_inout}
and \ref{stack_meandiff} reveals several striking properties
of the travel-time shifts, including frequency dependencies of
$\delta \tau_{\rm{+}}$, $\delta \tau_{\rm{-}}$, $\delta \tau_{\rm{mean}}$, and 
$\delta \tau_{\rm{diff}}$ at all phase-speed filters, and a surprising
connection between the sign of
the shifts and the value of central frequency of the filter 
with respect to the frequency of the $p_1$ ridge.
In particular, positive travel-time shifts (for both incoming
and outgoing waves) are observed exclusively in frequency
bandwidths that are centered below the $p_1$
ridge, shown by the solid line in Figures~\ref{stack_inout} and
\ref{stack_meandiff}. Maps of $\delta \tau_{\rm{+}}$, 
$\delta \tau_{\rm{-}}$, $\delta \tau_{\rm{mean}}$, and
$\delta \tau_{\rm{diff}}$ for filters with 
frequencies closest to the $p_1$ 
ridge, {\it i.e.} 4B, 3C, and 2D (not shown), show positive shifts 
near sunspot umbra and negative travel-time shifts elsewhere
in the ARs. At frequencies above (below) these 
values, the filters yield exclusively negative (positive)
travel-time shifts.
Filters E\,--\,J (not shown) exhibit trends 
similar to filter D. For these filters, negative values
for $\delta \tau_{\rm{+}}$ and  $\delta \tau_{\rm{-}}$ 
are observed throughout the
active region. Both incoming and outgoing time shifts 
measured with the larger phase-speed filters increase
with increasing frequency, with the values of the outgoing shifts exceeding
the incoming shifts.  These filters (E\,--\,J) show 
AR travel-time differences ($\delta \tau_{\rm{diff}}$) that decrease 
with increasing frequency.

\begin{figure}    
\centerline{\includegraphics[width=\textwidth,clip=]{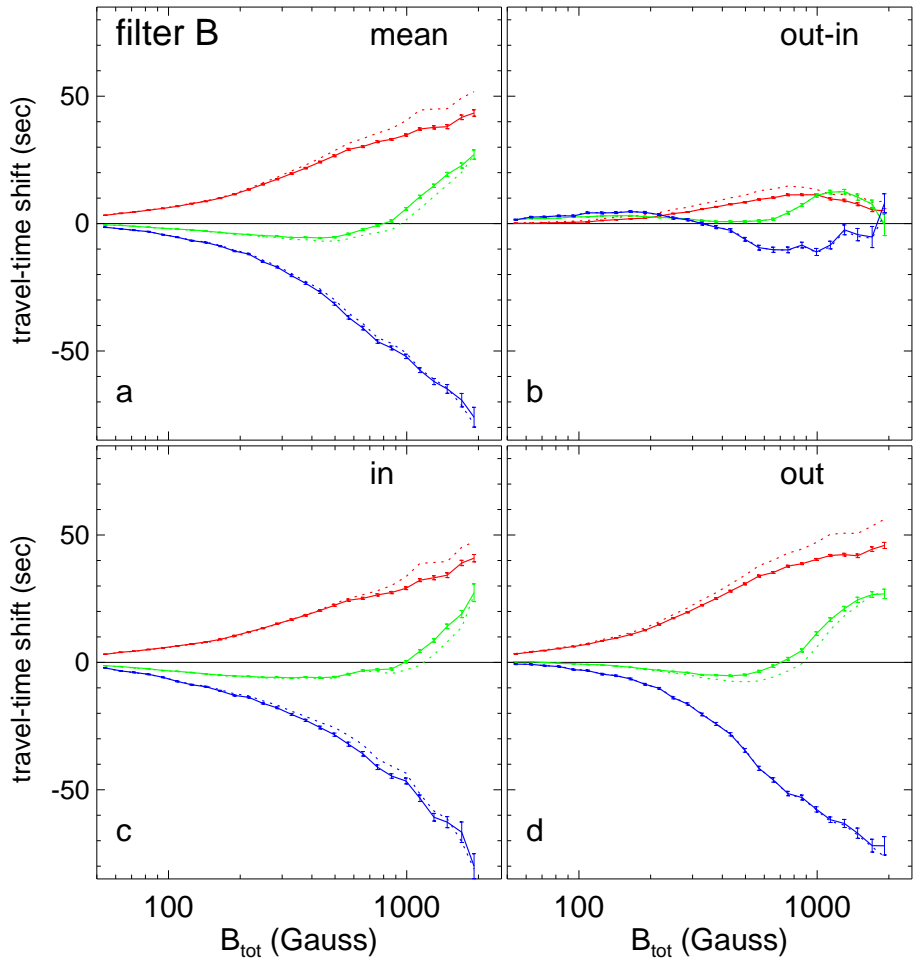}
}
\caption{
Travel-time shifts for filter B: 
a) the mean travel-time shift, b) the travel-time
difference, c) the incoming travel-time shift, and d) the
outgoing travel-time shift as functions of the 
magnetic field $B_{\rm tot}$. 
Solid (dotted) lines connect averaged
travel-time shifts, averaged over equally spaced bins
in the logarithm of the flux density, computed with (without) an 
amplitude-suppression correction (see text).  
Red, green and blue lines indicate frequencies
of 3,4, and 5 mHz respectively. 
Vertical bars indicate the total deviation (maxima minus minima) 
of the averages between three 
independent sub-regions containing the three largest sunspots.
}
\label{dt_vs_b_b}
\end{figure}

\begin{figure}    
\centerline{\includegraphics[width=\textwidth,clip=]{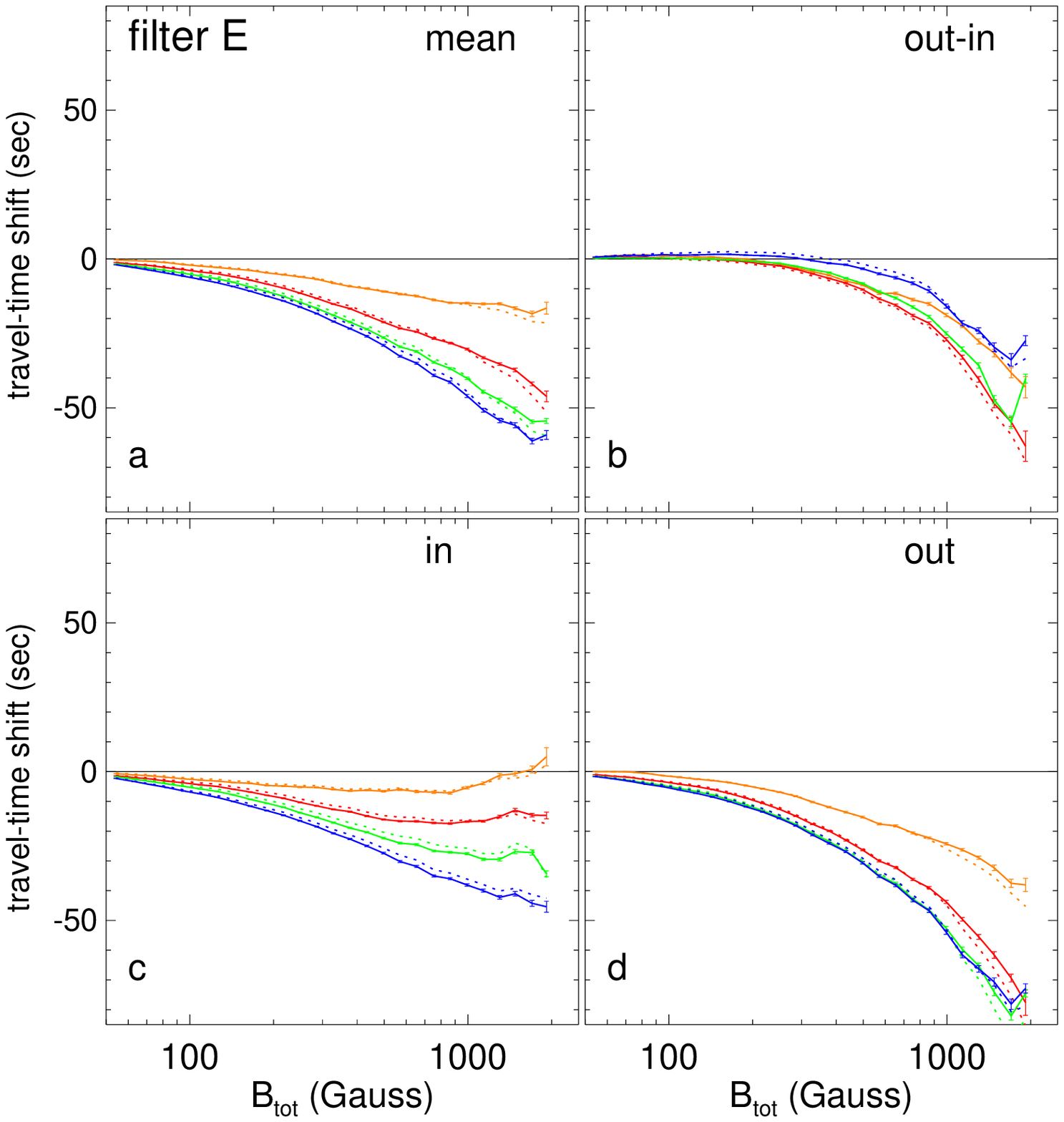}
}
\caption{
Travel-time shifts for filter E: 
a) the mean travel-time shift, b) the travel-time
difference, c) the incoming travel-time shift, and d) the
outgoing travel-time shift as functions of the 
magnetic field $B_{\rm tot}$. 
Solid (dotted) lines connect averaged
travel-time shifts, averaged over equally spaced bins
in the logarithm of the flux density, computed with (without) an 
amplitude-suppression correction (see text).  
Orange, red, green and blue lines indicate frequencies
of 2,3,4, and 5 mHz respectively. 
Vertical bars indicate the total deviation (maxima minus minima)
of the averages between three
independent sub-regions containing the three largest sunspots.
}
\label{dt_vs_b_e}
\end{figure}

As noted earlier by \inlinecite{Braun2006}, the
travel-times are non-linearly related to the
surface magnetic-flux density. The quantity $B_{\rm tot}$ is
derived from a MDI line-of-sight magnetogram assuming the magnetic field is
the gradient of a potential, and
used as a proxy for the total flux density. 
Figures~\ref{dt_vs_b_b} and
~\ref{dt_vs_b_e} show plots of travel-time shifts
against $B_{\rm tot}$ for phase-speed filters B and E respectively. 
For clarity, the scatter of individual pixel values is not shown
(although see Figure 1 of \opencite{Braun2006} for some examples). Instead, 
Figures~\ref{dt_vs_b_b} and \ref{dt_vs_b_e} 
show the average of the shifts derived from bins equally spaced in
the logarithm of the flux density. Solid (dotted) lines indicate
time shifts computed with (without) the amplitude-suppression
correction discussed in Section~\ref{S-Analysis}. The effect of 
the correction is to decrease the travel-time shifts, especially 
at lower phase-speeds (and smaller pupils) and 
in regions of high flux densities typical of the sunspot umbrae
and penumbrae, by amounts on the order of one to ten seconds.
Vertical bars in Figures~\ref{dt_vs_b_b} and
~\ref{dt_vs_b_e}, and in most of the other plots shown in
this paper, indicate the total deviation (maxima minus minima) 
of the averaged value as determined within three 
independent sub-regions containing the three largest sunspots.
Thus, the bars include contributions not only from sources
of random error, but also potentially systematic differences between 
sunspots. Typically, however, these deviations as a whole are
very small.

The frequency variation shown for filter B (Figure~\ref{dt_vs_b_b}) is 
typical of the results with smaller phase-speeds (and smaller pupils) 
which undergo the transition from positive time-shifts (at sub-$p_1$
frequencies) to negative time-shifts (at frequencies higher than
the $p_1$ ridge). The different frequencies shown in 
Figure~\ref{dt_vs_b_b}a exhibit the three types
of dependence on flux density of the mean travel-time shift described
by \inlinecite{Braun2006}. 
The travel-time differences (Figure ~\ref{dt_vs_b_b}b)
show similar trends except at the highest flux densities, where
the results for all frequencies approach zero. For filter E
(Figure~\ref{dt_vs_b_e}), it is observed that the frequency 
variations are in general larger for $\delta \tau_-$ 
shifts than for $\delta \tau_+$. As noted earlier, the difference
$\delta \tau_{\rm{diff}}$ for these larger phase
speeds show larger shifts at lower frequencies. This
trend is opposite that observed with the mean shifts 
$\delta \tau_{\rm{mean}}$. 

\begin{figure}    
\centerline{\includegraphics[width=\textwidth,clip=]{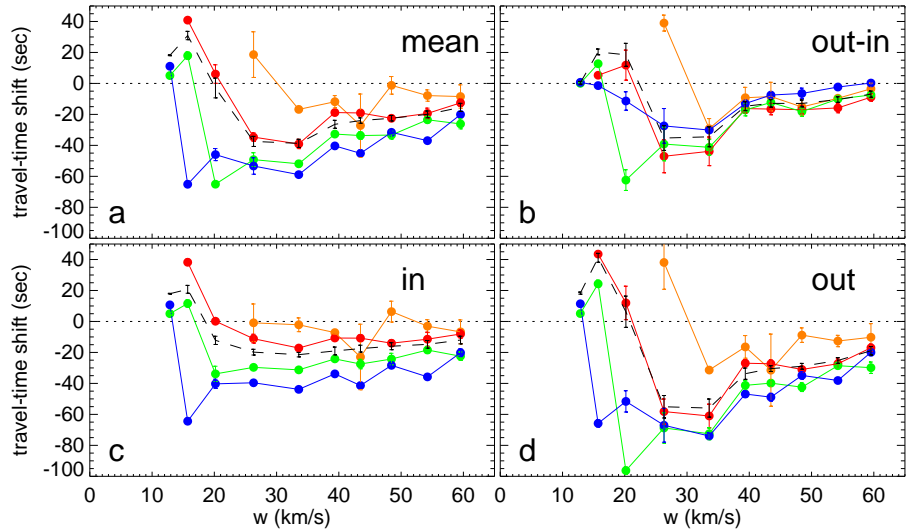}
}
\caption{a) The mean travel-time shift, b) the travel-time
difference, c) the incoming travel-time shift, and d) the
outgoing travel-time shift, averaged over the sunspot
umbrae, as functions of the phase speed $w$. 
Yellow, red, green and blue lines indicate frequencies
of 2,3,4, and 5 mHz respectively. The black dashed line
indicates the use of the wide (2.5\,--\,5.5 mHz) frequency bandpass.
Vertical bars indicate the total deviation (maxima minus minima) 
of the averages between three 
independent sub-regions containing the three largest sunspots.
}
\label{dt_vs_w_umb}
\end{figure}

\begin{figure}    
\centerline{\includegraphics[width=\textwidth,clip=]{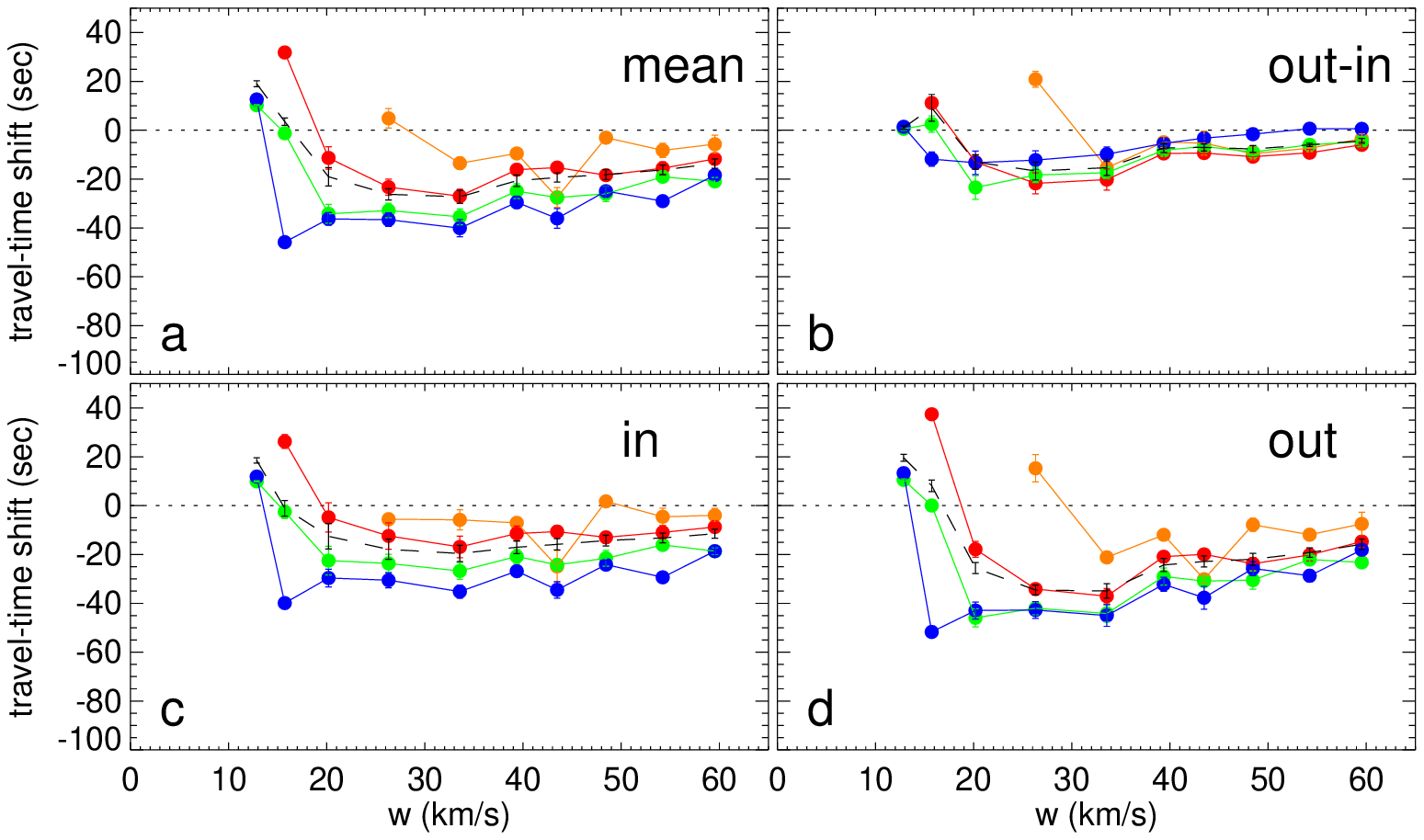}
}
\caption{
a) The mean travel-time shift, b) the travel-time
difference, c) the incoming travel-time shift, and d) the
outgoing travel-time shift, averaged over the sunspot
penumbrae, as functions of the phase speed $w$.
Yellow, red, green and blue lines indicate frequencies
of 2,3,4, and 5 mHz respectively. The black dashed line
indicates the use of the wide (2.5\,--\,5.5 mHz) frequency bandpass.
Vertical bars indicate the total deviation (maxima minus minima)
of the averages between three
independent sub-regions containing the three largest sunspots.
}
\label{dt_vs_w_pen}
\end{figure}

\begin{figure}    
\centerline{\includegraphics[width=\textwidth,clip=]{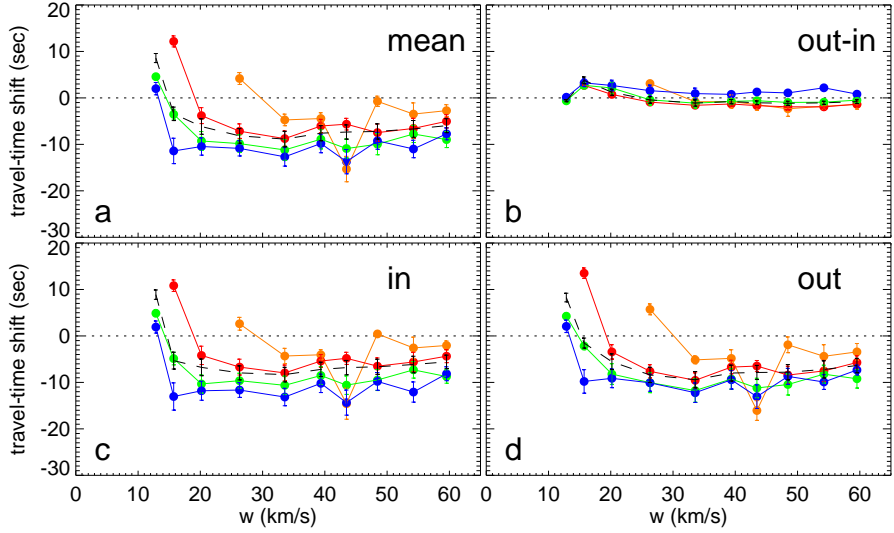}
}
\caption{
a) The mean travel-time shift, b) the travel-time
difference, c) the incoming travel-time shift, and d) the
outgoing travel-time shift, averaged over the 
plage, as functions of the phase speed $w$.
Yellow, red, green and blue lines indicate frequencies
of 2,3,4, and 5 mHz respectively. The black dashed line
indicates the use of the wide (2.5\,--\,5.5 mHz) frequency bandpass.
Vertical bars indicate the total deviation (maxima minus minima)
of the averages between three
independent sub-regions containing the three largest sunspots.
Note that the vertical scale differs from those in the figures for the umbrae
and penumbrae.
}
\label{dt_vs_w_plage}
\end{figure}

As \inlinecite{Braun2006} note, the close relationship 
between $\delta {\tau}_{\rm{mean}}$ and 
$B_{\rm tot}$ is consistent with
predominately near-surface perturbations, but does not rule
out subsurface perturbations that may very well correlate
with surface flux. As discussed in Section~\ref{S-Introduction}, 
the variation of travel-time shifts with both 
temporal frequency and phase speed is of 
critical importance in understanding the depth 
variations of the underlying perturbations. In particular,
\inlinecite{Braun2006} have suggested that the variation with frequency, 
at fixed phase speeds, of travel-time shifts may be a signature of 
surface effects (see also Paper 2).
To quantify these variations, we compute spatial averages of
the travel-time shifts over three types of
regions characteristic of the sunspot groups. 
The first two types are sunspot umbrae and penumbrae,
identified by brightness values less than 50\% and 92\% 
of the mean MDI continuum
values respectively. The third region of interest (which we
simply call ``plage'') is identified by values of
$B_{\rm{tot}}$ above 100 Gauss and excluding
areas previously marked as umbrae or penumbrae. 
The panel in the lowest-left corner of Figure~\ref{stack_meandiff}
illustrates in increasingly darker shades of grey the
three regions (plage, penumbrae, and umbrae) 
identified in this manner around NOAA 9885. 
The umbral and penumbral averages are of particular
importance since they represent time shifts
experienced by waves propagating through the immediate
subsurface layers of a sunspot (\eg within 10 Mm depth
below a 30 Mm diameter spot). Figures~\ref{dt_vs_w_umb},
~\ref{dt_vs_w_pen}, and ~\ref{dt_vs_w_plage} show the spatially
averaged time shifts for the umbrae, penumbrae and plage
respectively. These figures quantify many of the 
properties already noted in the travel-time shift maps;
including the strong frequency variations at each fixed
phase speed, and the changes of sign at smaller phase speeds.
Figures~\ref{dt_vs_w_umb}\,--\,\ref{dt_vs_w_plage} also include
the measurements made with the wide temporal bandpass (2.5\,--\,3.5 mHz).
These values, as expected from the
relative contributions of modes in the power spectra, fall
largely between the results obtained in the 3 and 4 mHz bandpasses.
Also noteworthy is that for the plage,
$\delta \tau_{\rm diff}$ varies from slightly negative at low
frequencies to slightly positive at high frequencies, and 
is mostly independent of phase speed.

It is noteworthy that essentially all of the travel-time
shifts observed in Figures~\ref{dt_vs_w_umb}\,--\,\ref{dt_vs_w_plage}
show significant frequency variations. At smaller phase speeds,
the variations of both mean travel-time shifts and travel-time
asymmetries show strong variations which often include a 
change of sign of the shifts. At higher phase speeds,
the mean travel-time shifts also show large
systematic frequency variations. The 
mean shifts observed at 5 mHz, for example,
are typically twice the value at 3 mHz, with the
difference being about 15\,--\,20 seconds. 

The effects of dispersion, including changes of 
ray paths as a function of temporal frequency
(\eg \opencite{Barnes2001}), may cause
frequency variations of travel-time shifts. 
However, the
frequency variations in the mean travel-time shifts observed here
may be much larger than are expected 
for sound-speed perturbations inferred from recent
inversions of travel-times.
Using the ray approximation, \inlinecite{Braun2006}
computed differences of the mean travel-time shifts between 
3 and 5 mHz, for a sound-speed perturbation similar to that 
of \inlinecite{Kosovichev2000}, of
about five to ten seconds for phase speeds
less than 30 km $s^{-1}$, and less than one to two seconds for 
higher phase speeds.
The implications of these variations are best explored
in the context of modeling (some initial efforts are
addressed in Paper 2). However, it is expected that observations
such as shown in Figures~\ref{dt_vs_w_umb}-\ref{dt_vs_w_plage}
may lead to methods for identifying and removing surface
effects. There are no precedents in either global
or local helioseismic inversions which offer any 
hope of including the contribution of
unresolved near-surface structure without making use
of the temporal frequency dependencies of the observables.


\subsection{Ridge Filters} 
  \label{S-ridge}

In this section, we present results obtained using filters
that isolate individual $p$-mode ridges. The use of
ridge filters allows us to judge
the sensitivity of the travel-time
shifts (especially those experienced by
waves near the $p_1$ ridge) to the choice of filter. 
Ridge-based filters also facilitate a more direct
comparison with results obtained with Fourier-Hankel
decomposition (see Section~\ref{S-hankel}).  Ridge filters have
been used previously in time-distance
helioseismology for $f$-mode studies
(\eg \opencite{Duvall2000}; \opencite{Gizon2002};
\opencite{Jackiewicz2007a}a; \citeyear{Jackiewicz2007b}b),
and recently for $p$ modes \cite{Jackiewicz2007c}.
Here we employ
filters for HH that isolate the $p_1$, $p_2$, $p_3$, and $p_4$ ridges.
At each temporal frequency, the filters have full 
transmission for wavenumbers spanning the midpoints between
the desired ridge and the neighboring ridges. Sharp
Gaussian roll-offs (similar to those used
to remove the $f$ mode in combination with the
phase-speed filters in Section~\ref{S-phasespeed}) remove the contributions 
above and below these wavenumbers. 

Unlike the common use of ray theory to define the radii of
annuli (TD) or pupils (HH), there 
is no ``standard'' procedure for adopting a pupil geometry
for ridge filters. After some trial and error, we 
settled on a fixed pupil for each ridge. We found
that the results were not largely dependent on the
outer pupil radius, but did change significantly with the
choice of inner pupil radius. A choice of inner pupil radius
smaller than roughly the horizontal $p$-mode wavelength of
the highest wavenumber present in the power spectra 
apparently produces
undesired leakage of the oscillatory signal at
the focus directly into the egression and ingression 
regressions. With these considerations, we adopted
a set of pupils with radii 9\,--\,42 Mm ($p_1$), 12\,--\,90 Mm ($p_2$),
14\,--\,167 Mm ($p_3$), and 17\,--\,195 Mm ($p_4$). We have also experimented
with varying the width of the frequency bandpass. The travel-time
shift maps analyzed here were made with $\Delta \nu = $ 0.26 mHz,
and were critically sampled with a frequency spacing of $\Delta \nu/2$. 

\begin{figure}    
\centerline{\includegraphics[width=0.75\textwidth,clip=]{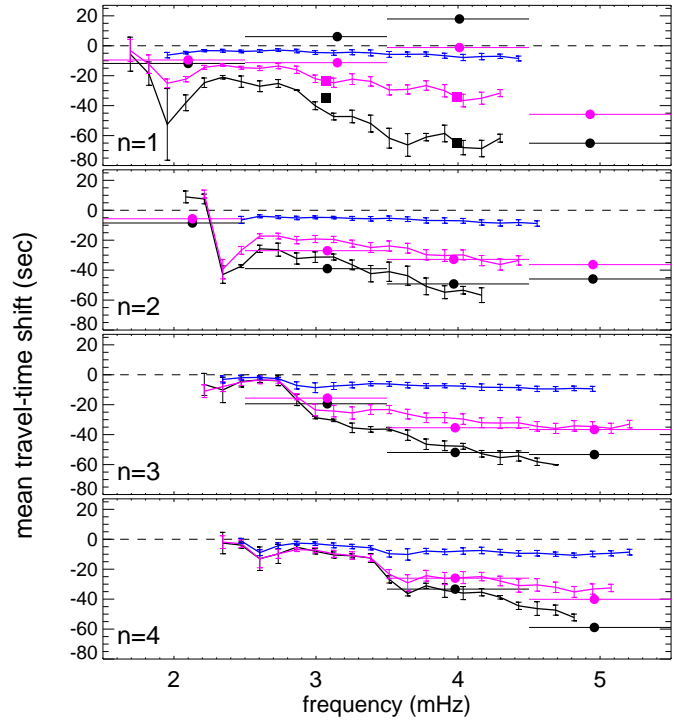}
}
\caption{Mean travel-time shifts
determined using  
ridge filters and averaged over umbrae (black lines), 
penumbrae (magenta lines), and plage (blue lines),
as functions of frequency along
the $p_1$ - $p_4$ ridges.
Vertical bars indicate the total deviation (maxima minus minima) 
of the averages between three 
independent sub-regions containing the three largest sunspots.
The filled black and magenta circles (with 1 mHz-wide horizontal
bars) represent comparisons of umbral and penumbral time-shifts 
determined from travel-time maps made with ``nearby'' phase-speed filters
(see text). The circles are placed at the power-weighted average
frequency for the given filter and frequency combination.
The squares shown near 3 and 4 mHz in the top panel 
indicate umbral (black)
and penumbral (magenta) mean travel-time shifts for the 
phase-speed filters (3D and 4C) which have 
higher values of phase-speed 
to those denoted by the filled circles (3C and 4B).
}
\label{tmean_ridge}
\end{figure}

\begin{figure}    
\centerline{\includegraphics[width=0.75\textwidth,clip=]{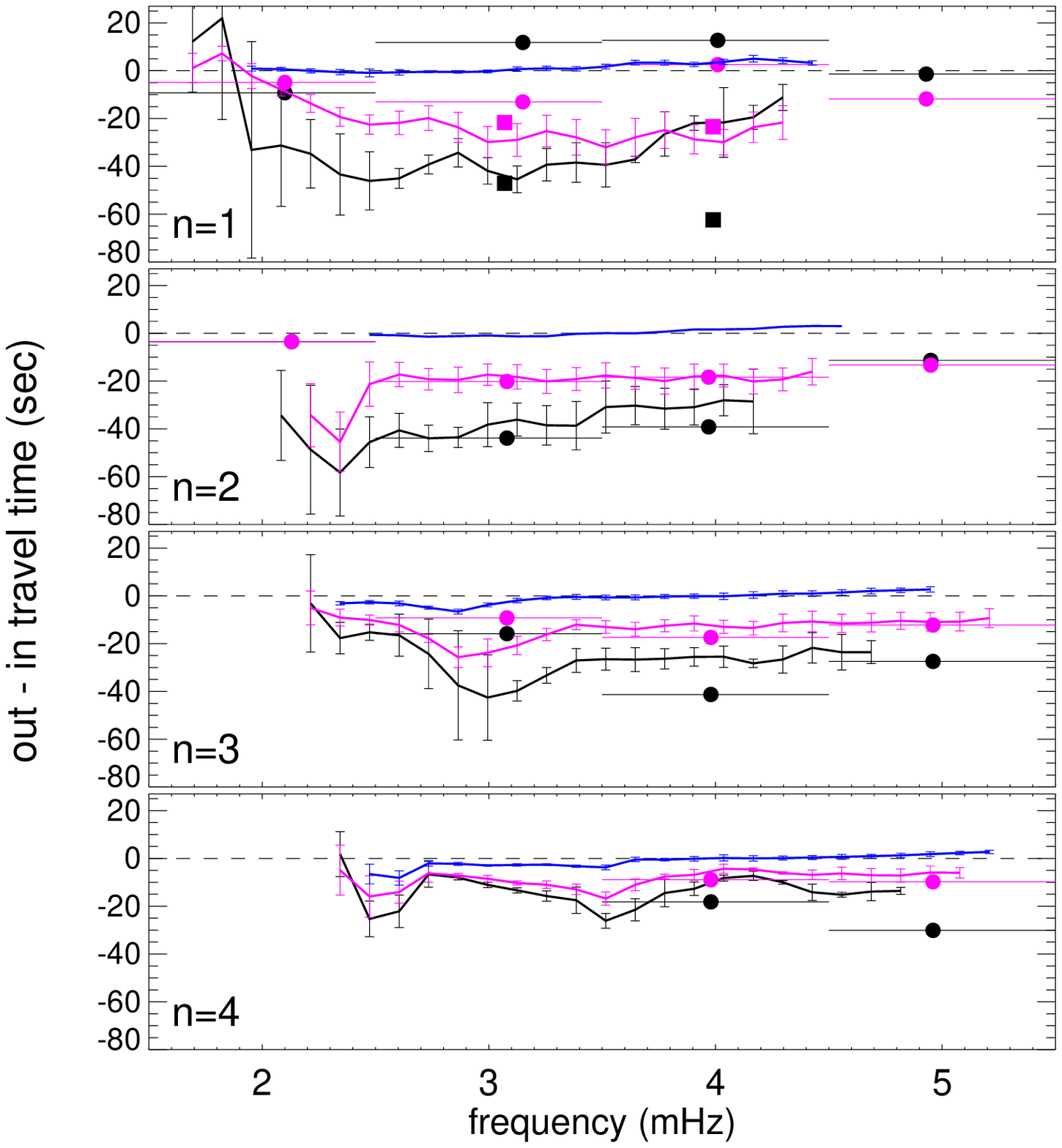}
}
\caption{Difference travel-time shifts determined using
ridge filters in the same format as Figure~\ref{tmean_ridge}
}
\label{tdiff_ridge}
\end{figure}

The results for the mean travel-time shifts, averaged over umbrae, penumbrae,
and plage, are shown in Figure~\ref{tmean_ridge}, while the results for
the travel-time differences are shown in Figure~\ref{tdiff_ridge}.
In light of the results using phase-speed filters (Section~\ref{S-phasespeed}),
what is most striking from Figures~\ref{tmean_ridge} and \ref{tdiff_ridge} is
that both the mean and difference travel-time shifts using ridge filters 
are essentially (with a few noisy exceptions) always negative. This is
also true along the $p_1$ ridge. Comparisons between the ridge-filtered
and phase-filtered results are facilitated by overlaying the nearest phase-speed-determined
values on Figures~\ref{tmean_ridge} and \ref{tdiff_ridge}. 
By ``nearest'' we mean that
for a given 1-mHz frequency bandpass and radial order, the nearest phase-speed filter
is that closest to the phase-speed of the ridge at
the central frequency. For $p_1$, the nearest filter
combinations are 2F, 3C, 4B, and 5B. For $p_2$, these combinations
are 2J, 3E, 4D, and 5C. For $p_3$, they are 3I, 4E, and 5D, and for $p_4$, they
are 4H and 5E.  For $p_2$, $p_3$, and $p_4$ there is very good agreement 
between phase-speed and ridge-filtered results 
for the mean travel-time shifts averaged in umbrae or penumbrae,
and reasonable agreement for the
time differences.  The largest discrepancies are clearly in the $p_1$
ridge, and (especially in the umbrae) involve 
a change in sign 
in the measurements of $\tau_{\rm mean}$ and $\tau_{\rm diff}$
between the two types of filters. 


It is noteworthy that the discrepancies between ridge-filtered
and phase-speed-filtered
results are largest for
phase-speed filters with frequency
bandwidths that are centered below $p_1$. 
These are the same phase-speed and frequency combinations 
which produce the positive
travel-time shifts seen in Figures~\ref{stack_inout} and \ref{stack_meandiff}.
In contrast, adjacent phase-speed filters centered above
the $p_1$ ridge apparently produce
travel-time shifts in sunspot umbrae and penumbrae ({\it i.e.}
the black and magenta squares in 
the top panels of Figures~\ref{tmean_ridge} and \ref{tdiff_ridge}),
which are very close to that observed with the $p_1$ ridge filter. 
Repeating the ridge-filtered measurements for some of
the cases where these sign changes occur (\eg near
filter 3C) with the same pupil as
used with the phase-speed filter yields yields deviations 
in both the mean shifts and travel-time asymmetry
of only a few seconds (out of a total of 30\,--\,40 seconds)
from results using the fixed pupil range stated earlier.
Thus, the discrepancy in the sign of travel-time perturbations between
the two types of filters is 
not the result of using different pupils. 

\begin{figure}    
\centerline{\includegraphics[width=0.75\textwidth,clip=]{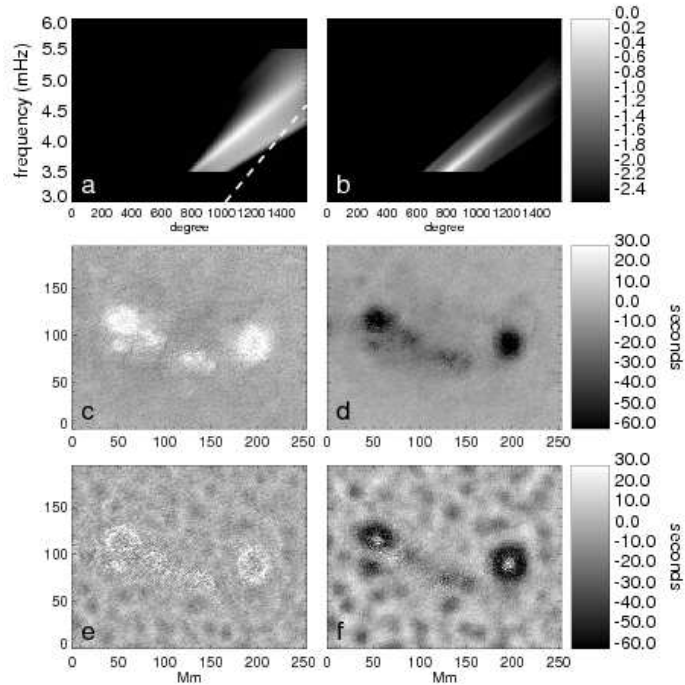}
}
\caption{ a) Azimuthally-averaged MDI power spectrum multiplied by
phase-speed filter ``1'' of Couvidat, \etal (2006). The grey
scale indicates the logarithm of the power, normalized to
the peak power. The dashed line indicates the mean phase speed of the
filter ($w = 12.8$ km $s^{-1}$). b) The
power spectrum at frequencies between 3.5 and 5.5 mHz multiplied by a 
$p_1$ ridge-filter, normalized to the peak power, 
c) mean travel-time shifts using the phase-speed filter,
d) mean shifts using the ridge filter, e) travel-time differences
using the phase-speed filter, and f) travel-time differences
using the ridge filter.
}
\label{filter}
\end{figure}

\begin{figure}    
\centerline{\includegraphics[width=0.75\textwidth,clip=]{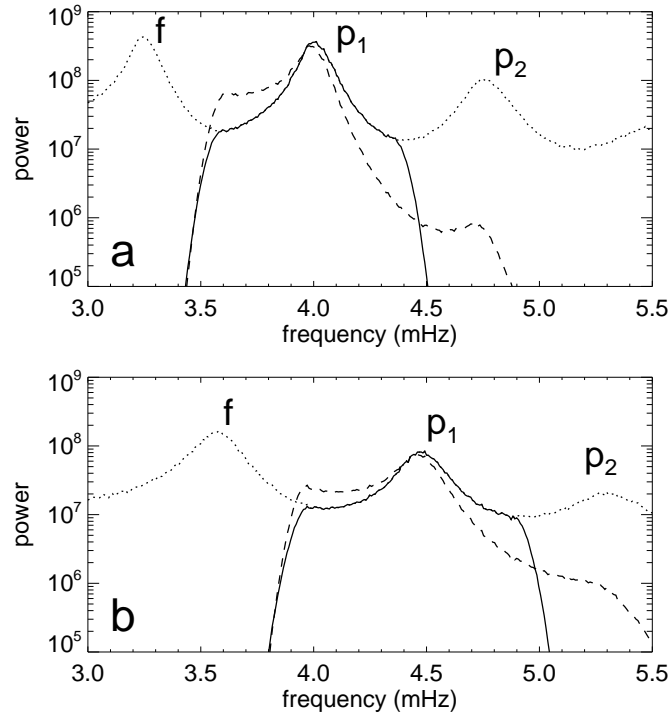}
}
\caption{ Cuts of the power as a function of frequency 
at two fixed wavenumbers
illustrating the effect of the filters shown in Figure~\ref{filter}:
a) power at a constant spatial wavenumber corresponding
to a spherical harmonic degree of 1055. The dotted line indicates
the unfiltered power, with the $f$, $p_1$, and $p_2$ ridges 
labeled. The solid line shows the power multiplied
by the ridge filter for $p_1$. The dotted line indicates the
power multiplied by a commonly used time-distance
phase speed filter (TD1; see text). For comparison with the
ridge-filtered power spectra, the phase-speed filtered
power spectrum is multiplied by a factor so that the integrated
power over the frequency bandwidth is the same as the ridge-filtered
spectrum. b) power as wavenumber  corresponding
to a spherical harmonic degree of 1230. For both cases, it
is clear that the principle difference between the two
filter types is the relative weighting of the two wings of
the $p_1$ ridge, such that the phase-speed filter enhances
the contribution of the low-frequency wing relative to the
high-frequency wing. 
}
\label{slices}
\end{figure}

Figure~\ref{filter} illustrates the extreme sensitivity of 
the sign of the travel-time shifts in sunspots near the $p_1$ ridge 
on the choice of filters. 
This figure compares the results of travel-time shifts computed 
with a commonly-used TD phase-speed filter (filter 1 of
\opencite{Couvidat2006} and \opencite{Zhao2006}; hereafter TD1) with shifts computed
with a $p_1$ ridge filter over a frequency bandpass between 3.5 and 5.5 mHz.
Despite the gross similarity of the filtered power included in the
measurements, the resulting maps
of $\tau_{\rm mean}$ and $\tau_{\rm diff}$ are drastically different.
The travel-time shift maps made with the TD1 filter
have larger positive values (\eg by about a factor of two in the mean travel-time
shift and a factor of four in the travel-time asymmetry in the penumbrae) 
than results obtained with the filter 5A
shown in  Figure~\ref{stack_inout},
even though the mean phase-speed of these filters are both approximately 13 km $s^{-1}$.
We have found that incrementally increasing the width of a 
phase-speed filter, centered at 12.8 km $s^{-1}$, produces maps with 
incrementally stronger positive travel-time shifts in sunspots.
Based on our experience with a variety of filters, we find in general 
that the requirement for producing positive travel-time shifts appears to be 
a disproportionate contribution to the correlations of wave power from 
the low-frequency wing of the $p_1$ ridge relative to the high-frequency
wing. This is illustrated in Figure~\ref{slices}.
This asymmetry apparently results from the fact 
that the mean phase speed of the filter (12.8 km $s^{-1}$; 
shown by the dashed line in Figure~\ref{filter}a) falls
significantly below the $p_1$ ridge.  

The reasons for the strong sensitivity, including
sign-changes, of the travel-time shifts 
to details of the filter (\eg width) and 
the relative weighting of the low-frequency wing of $p_1$ are
not fully understood at this time. 
\inlinecite{Thompson2007}
have presented evidence that a sign switch in mean travel-time shifts
also apparently occurs with the application of filters 
centered half-way ({\it i.e.} in the trough) between
the $p_1$ and $p_2$ ridges.

\subsection{Comparison with Fourier-Hankel Analysis} 
  \label{S-hankel}

\begin{figure}    
\centerline{\includegraphics[width=0.75\textwidth,clip=]{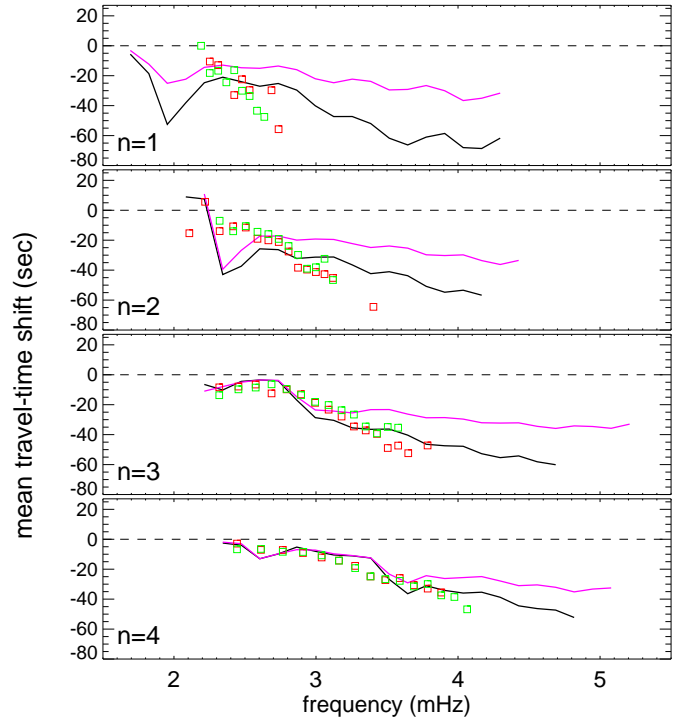}
}
\caption{The mean travel-time shift
determined using  
ridge filters and averaged over umbrae (black lines) and penumbrae 
(magenta lines) as shown in Figure~\ref{tmean_ridge}
as a function of frequency along
the $p_1$\,--\,$p_4$ ridges.  The squares
represent equivalent travel-time shifts
determined by previously published FH decomposition methods
applied to two sunspots, NOAA 5229 (red) and 5254(green).
}
\label{hankel}
\end{figure}

Here we compare the ridge-filtered (mean) travel-time shifts
with the phase-shifts observed in sunspots using Fourier-Hankel (FH) analysis 
\cite{Braun1995}. To do this, we use the published values of phase shifts
determined from Fourier-Hankel decomposition of waves 
around two sunspot groups, NOAA 5229 and 5254, observed with 
Ca {\sc ii} intensity images 
made at the geographic South Pole \cite{Braun1995}. Fortunately, the sunspots
studied by \inlinecite{Braun1995} are similar in size to those included in
this work. As noted in an earlier comparison
with TD measurements \cite{Braun1997}, 
the phase shifts in FH analysis
are divided by twice the angular frequency for comparison with
travel-time shifts (a switch in the sign of the Hankel results
is also needed due to different Fourier transform conventions). The results
are shown in Figure~\ref{hankel}.
The agreement between the two sets of measurements is very respectable, despite
differences in methods and data sets. In particular, the agreement is
significantly better than an earlier comparison
between FH phase shifts and TD travel-time shifts of 
\inlinecite{Braun1997}, which may be due to the poorer spatial resolution
and lack of mode discrimination ({\it i.e.} lack of filtering) in that
study.  The agreement is especially good for the $p_3$ and $p_4$ ridges,
and fair for the $p_2$ ridge. There are clear systematic differences
for $p_1$, however, with the FH results indicating stronger travel-time
shifts at higher frequencies than the HH results.

Even under ideal circumstances, the two methods (FH
decomposition and surface-focused HH) may be expected to yield systematic
differences. For example, surface-focused HH (like TD)
is primarily sensitive to the set of wave components that propagate
to the surface at chosen locations, unlike  FH decomposition which
provides no such discrimination. At low phase speeds, for example, phase shifts
and absorption coefficients determined from FH analysis  
include contributions from wave components passing under the sunspot and,
at high phase speeds, include contributions from waves which refract to the 
surface multiple times in the sampling annulus. The expectation is
that the results of
FH analysis may be less sensitive to perturbations at the target
than those obtained by using TD or HH.
In light of these considerations, the agreement shown in Figure~\ref{hankel}
is remarkable.  

\begin{figure}    
\centerline{\includegraphics[width=0.75\textwidth,clip=]{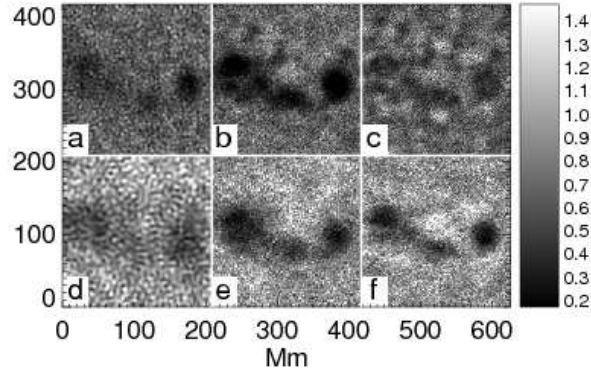}
}
\caption{Maps of the ratio of egression to ingression power
determined using  ridge filters and 0.26 mHz-wide frequency bandpasses centered at
a) 2.47 mHz, b) 3.52 mHz, and c) 4.55 mHz along the $p_1$ ridge, and 
d) 2.99 mHz, e) 4.04 mHz, and f) 5.08 mHz along the $p_3$ ridge.
At all frequencies, the areas containing sunspots (and other magnetic
flux) are dark, which corresponds to $\eta < 1$. 
At  high frequencies (panels c and f), however, note the additional
presence of brighter regions. 
} 
\label{abs_maps}
\end{figure}

\begin{figure}    
\centerline{\includegraphics[width=0.75\textwidth,clip=]{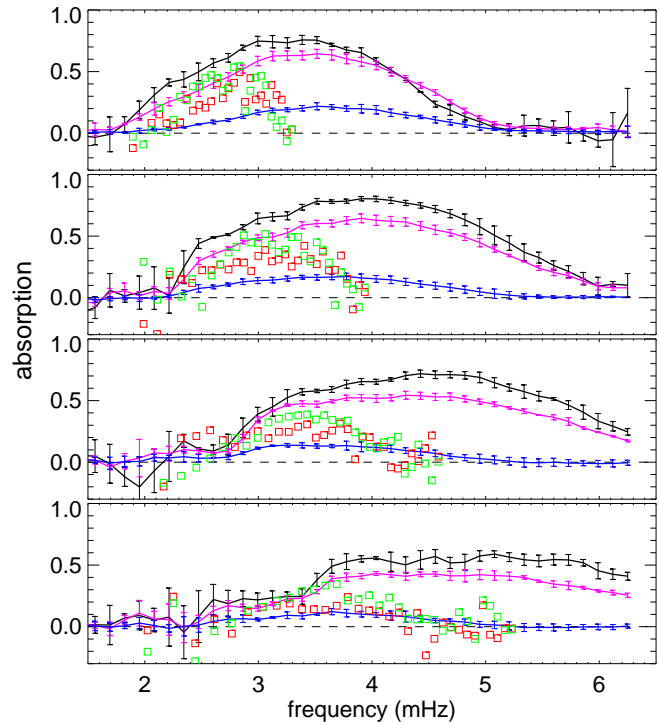}
}
\caption{The surface-focus HH absorption parameter $\alpha_{\rm HH}$ 
determined using  
ridge filters and averaged over umbrae (black lines) and penumbrae 
(magenta lines) as shown in Figure~\ref{tmean_ridge}
and plage (blue lines) as a function of frequency along
the $p_1$ - $p_4$ ridges.  The squares
represent values of the absorption coefficient 
determined by previously published Fourier-Hankel decomposition 
methods $\alpha_{\rm FH}$ applied to two sunspots, NOAA 5229 (red) and 5254 (green).
}
\label{abs_hankel}
\end{figure}

Maps of the ratio of egression to ingression power 
($| H_+ |^2 / | H_-|^2$) can be used in surface-focused HH to probe local
emission and absorption properties. Representative maps of this quantity,
computed from several frequency
bandpasses using the $p_1$ (top panels) and $p_3$ (bottom panels) ridge filters,
are shown in Figure~\ref{abs_maps}. 
The quantity $\alpha_{\rm HH} = 1 - | H_+ |^2 / | H_-|^2$ 
(which we denote as the HH absorption parameter) 
can be directly compared with
the absorption coefficients determined using FH
decomposition measurements, $\alpha_{\rm FH}$ (Figure~\ref{abs_hankel}).
The FH absorption coefficients
are typically smaller than the surface-focused HH absorption parameter
in either the umbrae or penumbrae. Peak values of $\alpha_{\rm HH}$ of
around 0.7 are observed in sunspot umbrae as compared to typical
peak values of $\alpha_{\rm FH}$ of around 0.5.
Also striking is the difference in behavior of $\alpha_{\rm FH}$ and 
$\alpha_{\rm HH}$ at high frequencies, with the absorption parameter
from FH methods decreasing
towards zero at much lower frequencies then the absorption 
parameter determined from HH. Some prior HH measurements
have shown evidence for this behavior \cite{Lindsey1999}, 
and it has been speculated that the presence of surrounding emission
(called ``acoustic glories''), which may not be readily
resolved by FH decomposition methods, 
may be at least partly responsible
for the lower observed values of $\alpha_{\rm FH}$.  It is highly
likely that the high-frequency (\eg $\nu \geq$ 5 mHz)
properties of the measured absorption parameters are determined not
only by actual absorption mechanisms in the sunspots but by
local emission properties as well. For example, a decrease
in the local emission can in principle decrease the egression power, and
hence the HH absorption parameter. What is known from measurements of
$p$-mode lifetimes suggests that such a mechanism, as an explanation for
all apparent absorption, is not viable for waves at lower frequencies,
where the contribution of acoustic
flux actually originating at a given target 
is expected to be only a very small
fraction of the total observed egression (or outgoing) power \cite{Braun1987}.
Finally, we urge caution regarding any interpretation of
the observed fall-off of $\alpha_{\rm HH}$ shown in Figure~\ref{abs_hankel}.
No attempt has been made to assess (and remove) 
any ``background'' contributions
to the measured egression and ingression (\eg contributions due to locally
generated oscillatory motion occurring within the  
pupils over which the ingression and egression amplitudes are evaluated).
Thus, the observed decrease towards zero of the HH absorption 
measurements may be unphysical.

\section{Discussion}
  \label{S-discussion}

It is our preference to let most of the results, as presented
through Figures~\ref{stack_inout}\,--\,\ref{abs_hankel}, speak
for themselves.  A summary of these observations
would likely be either too lengthy or otherwise incomplete in that
potentially important relationships not directly addressed 
the text may be neglected.  It ought to be fairly clear from observations
such as these, however, that the interaction between solar
magnetic regions and acoustic waves is highly complex, and that
no existing model is sufficient in explaining or predicting the
complete range of observed behavior. In addition, it should be kept
in mind that our observations represent measurements of only three active
regions, with only one observable, using one spectral line, and along 
(essentially) one line-of-sight. It is known, from other observations,
that dependencies of phase or travel-times shifts on these and other 
variables are now recognized, if not fully understood. Deep-focus
methods (as applied to TD or HH analyses) are expected
to provide further important constraints on models. We also note
that we have only briefly touched on the observations relevant
to $p$-mode absorption in magnetic fields, with the expectation that we
will return to this in further publications. 

An open question is the degree to which
``surface effects,'' 
unknown or unaccounted for physical influences of magnetic fields
on acoustic waves, are important in the modeling of 
subsurface structure of sunspots and active regions.
As the set of observations shown here confirm and
expand upon those presented by \inlinecite{Braun2006}, it is worth
restating the general conclusions derived there. Namely,
the strong frequency variation of the measured travel-time
shifts cannot be explained using standard assumptions,
{\it i.e.}\ standard ray-approximation based modeling applied
to sound-speed models that are typical of
published 3D inversion results. To these unexplained
frequency variations must
now be added apparently ``anomalous'' positive travel-time
shifts (both the mean and travel-time asymmetry) in sunspots, 
so called here because the conditions ({\it i.e.} $p$-mode 
properties and choice of filter),
under which they appear defy the expectations
of standard assumptions and models. We will return to
both of these issues, in the context of models of sound-speed 
perturbations, in Paper 2.   

It has been argued (\eg \opencite{Zhao2006}) that
the sign-change of travel-time shifts with varying
phase speed provides
evidence for the relative lack of importance of 
surface effects for standard inferences
from 3D inversions. Certainly, a change of sign
is not a typical property of known ``surface terms'' 
in models of frequency shifts in structural
inversions in global helioseismology and 
ring-diagram analyses. However, rather than identify
the sensitivity of the sign of
$\tau_{\rm mean}$ or $\tau_{\rm diff}$ to the choice of filter
with a magnetic surface effect, 
these observations seriously raise the possibility that 
the positive values of these shifts represent an artifact,
by which we mean a property that
is more sensitive to the methods of the analysis than to
actual physical conditions within or below sunspots.

\begin{acks}
We thank an anonymous referee for useful suggestions.
This work is supported by funding through NASA contracts 
NNH04CC05C, NNH05CC76C, NNG07E151C, NSF grant AST-0406225, 
and a subcontract through the
HMI project at Stanford University awarded to NWRA.
\end{acks}

\bibliographystyle{spr-mp-sola}
\bibliography{db}

\end{article} 
\end{document}